\documentstyle[epsf,times]{aa}

\def\rerg{\rm erg} 
\def\rs{\rm s} 
\def\rs1{\rm s$^{-1}$}

\def\rcm{\rm cm} 
\def\rcm2{$\rm cm^{-2}$} 
 
\def\ergs{\rerg\ \rcm2\ \rs1} 
\def\flux{\mbox{ } {\rm erg} \mbox{  } {\rm cm}^{-2}}

\def\etal{{et al.}} 
\def\apj{{ApJ }\ } 
\def\apjl{{ApJ }\ }

\def\nature{{Nat }\ } 
\def\aa{{A\&A}\ } 
\def\aas{{A\&AS}\ } 
\def\mnras{{MNRAS}\ } 
\def\iauc{{IAU Circ.}\ }
  
\thesaurus{03(13.07.1; 13.07.2; 13.25.3)}

\title{BeppoSAX spectrum of GRB971214: evidence of a substantial energy
output during afterglow}
  
\author{D. Dal~Fiume\inst{1}
 \and L. Amati \inst{1} \and L. A. Antonelli\inst{2,3} \and F. Fiore\inst{2,3}
 \and J. M. Muller\inst{2,4} \and A. Parmar\inst{5} \and N. Masetti\inst{1}
 \and E. Pian\inst{1}
 \and E. Costa\inst{6} \and F. Frontera \inst{1,7} \and L. Piro\inst{6}
 \and J. Heise\inst{4} \and R. C. Butler\inst{8} \and A. Coletta\inst{3}
 \and M. Feroci\inst{6} \and P. Giommi\inst{2}
 \and L. Nicastro\inst{9} \and  M. Orlandini\inst{1}
 \and E. Palazzi\inst{1} \and G. Pizzichini\inst{1} 
 \and M. Tavani\inst{10}}
  
\institute{
Istituto di Tecnologie e Studio delle Radiazioni Extraterrestri
 (TeSRE), C.N.R., via Gobetti 101, I-40129 Bologna, Italy
\and Beppo-SAX Scientific Data Center, Via Corcolle 19, I-00131 
 Roma, Italy 
 \and
 Osservatorio Astronomico di Roma, Via Frascati 33, I-00044 Roma, Italy
 \and
 Space Research Organization Netherlands, 
 Sorbonnelaan 2, NL-3584 CA Utrecht, The Netherlands
 \and
 Astrophysics Division, Space Science Department of ESA, 
 ESTEC, P.O. Box 299, NL-2200 AG Noordwijk, The Netherlands
 \and
 Istituto di Astrofisica Spaziale (IAS), C.N.R., Via Fosso del Cavaliere,
 I-00133 Roma, Italy
 \and
 Dipartimento di Fisica, Universit\`a di Ferrara, Via Paradiso 12,
 I-44100 Ferrara, Italy
 \and
 Agenzia Spaziale Italiana,  Viale Regina Margherita 202, 
 I-00198 Roma, Italy
 \and
 Istituto di Fisica Cosmica con Applicazioni all'Informatica (IFCAI),
 C.N.R., Via U. La Malfa 153, I-90139, Palermo, Italy
 \and
 Istituto di Fisica Cosmica ``G. P. S. Occhialini'',
 C.N.R., via Bassini 15, I-20133 Milano, Italy
 }
\offprints{D.~Dal~Fiume --- daniele@tesre.bo.cnr.it}
\date{Received [date]; accepted [date]}

\begin{document} 

\titlerunning{GRB971214 and its afterglow}
\authorrunning{D. Dal Fiume et al.}
\maketitle

\begin{abstract}
We report the X/$\gamma$-ray spectrum of GRB971214 and of its afterglow.
The afterglow was
measured few hours after the main event and for an elapsed time of more
than two days. The measure of this GRB and afterglow is relevant due to
its extreme, cosmological distance (z=3.42). The prompt event shows a
hard photon spectrum, consistent with a broken power law with photon
indices $\Gamma_{\rm X}\approx$0.1 below $\sim$20 keV and
$\Gamma_\gamma\approx$1.3 above 60 keV.
The afterglow spectrum, measured with the MECS and LECS
BeppoSAX telescopes, is consistent with a power law with spectral photon
index $\Gamma$=1.6. Within the statistical accuracy of our measure no
spectral evolution is detected during the observation of the afterglow.
When integrated during the time span covered by BeppoSAX
observations, the power in the afterglow emission, even with very
conservative assumptions, is at least comparable with the power in the
main event. The IR-to-X rays broad band spectrum is also presented,
collecting data from the literature and adding them to the BeppoSAX measure.
It shows that the predictions from synchrotron emission models is
qualitatively confirmed. The BeppoSAX measurement of the X and $\gamma$
ray spectrum of this GRB/afterglow is discussed in the framework of
current theoretical models
\end{abstract}


\section{Introduction}

The discovery of X-ray afterglows from Gamma-ray Bursts (GRBs) (Costa et
al. \cite{costa})
is a major step forward to understand this still mysterious
phenomenon. The detection of the faint, fading X-ray counterparts of GRBs
poses tight constraints to the models for the emission. Multiwavelength
studies discovered optical, IR and radio transients associated with the
X-ray afterglow, thanks to the unprecedented positioning accuracy
obtained with BeppoSAX. The discovery of a substantial redshift
(Kulkarni et al. \cite{kulkarni_c,kulkarni_n}) in the absorption 
and emission lines in the spectra of the host galaxies associated
with the GRBs optical transients puts these catastrophic events at a
cosmological distance and results in an extreme energy output from each
GRB, if the emission is isotropic. 

The recent advances in our knowledge about the cosmic events
known as GRBs were mainly due to the accurate positioning allowed by
BeppoSAX (Boella et al. \cite{boella}).
This satellite carries on board an optimal set of
instruments to detect GRBs (the Gamma Ray Burst Monitor - GRBM
Frontera et al. \cite{frontera}, Feroci et al.  \cite{ferocigrbm} ),
to position them within a few arcminutes (the Wide
Field Cameras - WFC; Jager et al. \cite{jager})
and finally to pinpoint the positions
down to tens of arcseconds thanks to rapid (few hours) follow-up 
observations with the
Narrow Field Instruments (Low Energy Concentrator Spectrometer - LECS;
Parmar et al. \cite{parmar};
Medium Energy Concentrator Spectrometer - MECS; Boella et al.
\cite{boella2}; High Pressure Gas Scintillation Proportional Counter -
HPGSPC; Manzo et al. \cite{manzo};
Phoswich Detection System - PDS; Frontera et al. \cite{frontera} )

The positions given by BeppoSAX (e.g. Piro et al. \cite{piro} for
\object{GRB960720}) allowed prompt ground based observations
with telescopes in optical, radio, IR. Up to now thirteen Optical
Transients (OT) were discovered in the error boxes of the X-ray
afterglows (van Paradijs et al.  \cite{ot_1}, Bond \cite{ot_2},
Halpern et al.  \cite{halpern}, Groot et al. \cite{ot_3}, Palazzi et al.
\cite{ot_4}, Galama et al. \cite{ot_5}, Jaunsen et al. \cite{ot_6},
Hjorth et al. \cite{ot_7}, Bloom et al. \cite{ot_8},
Kulkarni et al. \cite{ot_9}, Galama et al. \cite{ot_10}, Palazzi et al.
\cite{ot_11}, Bakos et al. \cite{ot_12}). 
After the fading of the OT in most cases a faint galaxy was detected.
The detection of the putative host galaxy of GRB971214 is particularly
intriguing as the estimate of the
redshift is z=3.42, locating this event at an extreme cosmological
distance (Kulkarni et al. \cite{kulkarni_n}). 

We observed with BeppoSAX a GRB on December 14.97272 UT 1997
(Heise et al. \cite{heise_grb}). A follow-up pointing performed 6.67 hours
after the main event detected the faint and fading X-ray source
\object{1SAX J1156.4+6513}
(Antonelli et al. \cite{antonelli}). After the fading
of the optical transient, spectroscopic observations of the associated
host galaxy tentatively put it at cosmological distance, as the estimate of
its redshift is z=3.42 (Kulkarni et al. \cite{kulkarni_n}),
that corresponds to a luminosity distance $>$30 Gpc (for H$_0$=65 km s$^{-1}$ 
Mpc$^{-1}$ and $\Omega_0$=0.2). The
complete evolutionary history of the emitted spectrum from X to $\gamma$
rays up to 2.5 days after the main event suggests that the afterglow in
X-rays begins immediately. The energy output in the afterglow results to
be comparable to that in the prompt event.
With the new wealth of data on optical counterparts and X--ray
afterglows, a major revamping of theoretical models is
occurring. With the extragalactic origin firmly established on the basis
of observations of host galaxies (Metzger et al. \cite{metzger},
Kulkarni et al. \cite{kulkarni_n, ot_9}), the cosmological fireball model
(e.g. Cavallo \& Rees \cite{cavallo}, M\'esz\'aros \& Rees
\cite{meszrees1}) gives predictions that reasonably fit to
observational data. In this paper we discuss
the details of the prompt and delayed emission from GRB971214, with
emphasis on the X and $\gamma$ ray spectrum and on its evolution with time.
Implications on the models of the prompt event and of the afterglow
are discussed.

\section{Observations}
\object{GRB971214} was detected in Lateral Shield 1 of GRBM and in WFC 1 on
December 14.97272 UT. The burst lasted approximately 30 s, with two
leading broad peaks (3s and 10 s FWHM) and a third fainter and sharper
peak (1s FWHM) 34s after trigger. The GRB profile is shown in Fig.
1, as measured by GRBM LS1 and WFC1.

\begin{figure}
\epsfxsize=\columnwidth
\epsfbox{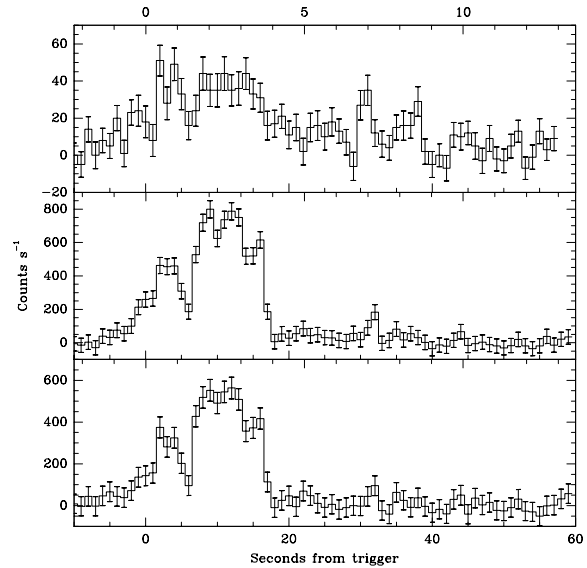}
\caption[]{
GRB971214 as observed by BeppoSAX. Top panel: WFC1 (1 second time
resolution -- E=1--26 keV).
Middle panel: GRBM LS1 ratemeter (1 second time resolution -- E=40--700 keV).
Bottom panel: GRBM LS1 AntiCoincidence ratemeter (1 second time
resolution -- E$>$100 keV).
Zero time corresponds to the GRBM BeppoSAX trigger time.
Top scale: time in the GRB reference frame (assuming z=3.42). Bottom
scale: time in the Earth reference frame.}
\label{fig:lcurve}
\end{figure}

The analysis of the WFC data was performed using WFC
data reduction software version 103106.
Data reduction, background subtraction
and spectral analysis of GRBM data were performed using dedicated SW
tools (Amati et al. \cite{amati})
The latest release of the WFC and GRBM response matrices were used.
All the spectral fits reported in this article were performed
using XSPEC, version 10.0 (Arnaud \cite{xspec}).

\begin{figure}
\epsfxsize=\columnwidth
\epsfbox{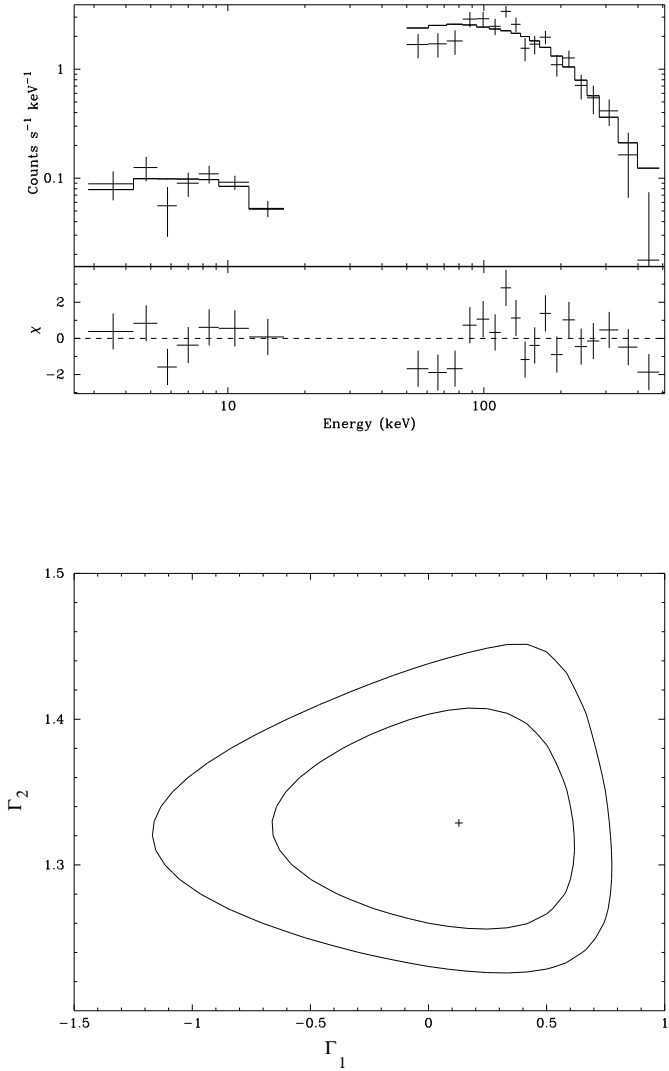}
\caption[]{
Top panel: the emitted spectrum of GRB971214 in the $\gamma$-ray 
and X-ray band. A fit with a broken power law is indicated as a 
solid line. Bottom panel: the joint confidence contour for $\Gamma_1$
and $\Gamma_2$. 68\% and 90\% confidence levels are shown.
}
\label{fig:grbspectrum}
\end{figure}

We fitted the joint spectrum with the spectral shape of Band et al.
(\cite{band})
($\chi^2_{\rm dof}$=1.99 for 22 degrees of freedom) and with a power law 
with exponential cutoff ($\chi^2_{\rm dof}$=1.98 for 22 degrees of freedom).
Both give systematic residuals in the WFC energy band.
The best fit is obtained using a broken power law ($\chi^2_{\rm dof}$=1.6
for 21 degrees of freedom). An F--test shows that the improvement is not
significant.
The average joint Wide Field Camera/Gamma Ray Burst Monitor spectrum 
during the burst can be well described by a broken power
law with photon indices $\Gamma_1$=0.13 and $\Gamma_2$=1.3 and a break energy 
at $\sim$10 keV. Given the gap between WFC and GRBM spectra,
approximately between 10 and 50 keV, a large uncertainty in the position
of the break is present and must be added to the statistical uncertainty
quoted in Table 1.
The fits with all the above functions are statistically unacceptable,
but we estimate that
a substantial contribution to the high value of $\chi^2_{\rm dof}$ is
due to systematic uncertainties in the spectral deconvolution. 
Therefore we do not consider to add further components to improve the fit.
The results from all the spectral fits we performed are reported in Table 1.
In Fig. 2 we report the count rate spectrum from WFC1 plus GRBM LS1.
In the same figure we report the joint confidence contours for the two
spectral indices of the broken power law.

\begin{table*}
\caption{Spectral fit parameters}
\begin{flushleft}
\begin{tabular}{llllll}
\noalign{\smallskip}
\hline
\noalign{\smallskip}
Start$^{(1)}$  &   End$^{(1)}$  &  $\Gamma_1$  &  $\Gamma_2$  &  
E$_{\mbox{cut}}$ & $\chi^2_{\rm dof}$ \\
time   &   time &  &  & (keV) & (dof)\\
\noalign{\smallskip}
\hline
\noalign{\smallskip}
\multicolumn{6}{c}{Average spectra}\\
\noalign{\smallskip}
\hline
\noalign{\smallskip}
0 & 35&0.13$^{+0.33}_{-0.47}$&1.33$\pm$0.05&10$^{+2.3}_{-1.5}$&1.6(21)\\
23300 & 215000 &1.6$\pm 0.2$& & &0.94(20)\\
\noalign{\smallskip}
\hline
\noalign{\smallskip}
\multicolumn{6}{c}{Time resolved spectra}\\
\hline
\noalign{\smallskip}
0 & 9 &0.37$\pm$0.23&1.4$\pm$0.11&95$\pm$14&0.64(12)\\
9 & 20&0.33$\pm$0.27&1.0$\pm$0.02&50$\pm$4&0.63(12)\\
23300 &75000&1.9$^{+0.39}_{-0.34}$& & &0.7(11)\\
75000 &125000&1.42$^{+0.85}_{-0.68}$& & &0.28(5)\\
125000&215000&1.23$^{+0.75}_{-0.91}$& & &0.44(2)\\
\noalign{\smallskip}
\hline
\noalign{\smallskip}
\multicolumn{6}{l}{\footnotesize (1) Seconds from trigger time}\\
\multicolumn{6}{l}{\footnotesize NOTE: errors are single parameter 68\%
confidence level}
\end{tabular}
\label{tab:spectralfits}
\end{flushleft}
\end{table*}

The joint confidence contour between $\Gamma_2$ and the break energy
(not reported in Fig. 2)
suggests that the GRB spectrum shows a bending above $\approx$ 10 keV but
the hard X-ray/$\gamma$ ray spectrum above 50 keV is consistent with a
single power law. A fit with a power law plus exponential cutoff gives
an e-folding energy of $\sim$300 keV.
This is in the "hard" tail of the distribution from statistical studies of 
BATSE GRB spectra (Band et al. \cite{band}),
in which a broad interval of break energies from 50
keV to more than 1 MeV was found, with the majority of GRBs having a
cutoff energy below 200 keV, and confirms that this GRB has a very hard
spectrum.

We analyzed separately the spectra of the first and of the second peak.
We find evidence that the second peak is at least as hard as the first one.
This result is different from the hard-to-soft evolution that
seems to be present in most GRBs (Preece et al. \cite{preece}).
In this respect
the properties of GRB971214 are distinct from those observed on the
average in GRBs.

The burst fluences are 1.9
$\pm 0.4 \times 10^{-7} \flux$ in 2-10 keV and 8.8 
$\pm 0.8 \times 10^{-6} \flux$ in 40-700 keV.
The fluence in hard X-rays/$\gamma$ rays is in good agreement with 
that measured with BATSE (Kippen et al. \cite{batse_fluence}).
Assuming a redshift z=3.42 (Kulkarni et al. \cite{kulkarni_n})
in a standard Friedmann
cosmology (with H$_0$=65 km s$^{-1}$ Mpc$^{-1}$ and $\Omega_0$=0.2) the
luminosity distance is 1.05$\times 10^{29}$ cm. At this distance the
observed fluences correspond to 6 $\pm 1.2 \times 10^{51} $  ergs in
2-10 keV and 2.8$\pm 0.25 \times 10^{53} $ ergs in 40-700 keV for an
isotropically emitting source. Here we assume that
L$_{\rm grb}={\rm F}_{\oplus} 4\pi {\rm D}_{\rm L}^2 (1+{\rm z})^{-1}$
(e. g. Hakkila et al. \cite{hakkila}).
If we assume the measured slope of the GRB spectrum (see Table 1),
L$_{\rm grb}={\rm F}_{\oplus} 4\pi {\rm D}_{\rm L}^2 (1+{\rm z})^{-1.3}$
and the luminosity is a factor 1.6 lower in the hard band.
The energy ranges correspond to 4.4-44
keV and 180-3090 keV at the source. The X-to-$\gamma$ fraction
is therefore 0.02. This can be compared to other fractions measured
for other GRBs as reported in Frontera et al. (\cite{frontera_grb})
that range from 0.39 for \object{GRB970508} to 0.01 for 
\object{GRB980329}. Therefore this GRB
shows one of the lowest ratios, i. e. the hardest spectrum, amongst those
observed with BeppoSAX. Of course such a comparison is made {\it
without} taking into account possible substantial differences in redshift
amongst the different GRBs.
If the emitted X-$\gamma$ ray spectrum has a break somewhere above 10
keV, the redshift due to the extreme cosmological distance shifts this
break to a lower energy in the spectrum as observed at earth, possibly
affecting the fluence ratio.

\begin{figure}
\epsfxsize=\columnwidth
\epsfbox{grb971214.f3}
\caption[]{
Top panel: the emitted spectrum of 1SAX J1156.4+6513 in the 
0.7-10 keV energy band as measured with the LECS and MECS telescopes.
Bottom panel: the joint confidence contour for $\Gamma$ and N$_{\mbox{H}}$.
68\% and 90\% confidence levels are shown.\\
Note that the model assumes a redshift {\it z}=3.42.}
\end{figure}

After the detection of GRB971214 and its positioning using WFC1
(Heise et al. \cite{heise_grb}),
BeppoSAX was rescheduled to point the center of the WFC error box.
The observation started on December 15.24583 UT ($\sim$6.5 hours after the
gamma--ray burst) and lasted until December 17.50069 UT for a total
elapsed time of 2.25 days.
A faint source, 1SAX J1156.4+6513 at $\alpha_{2000}$=
11$^{\rm h}$ 56$^{\rm m}$ 25$^{\rm s}$ and 
$\delta_{2000}$=+65$^{\rm o}$ 13' 11'' (Antonelli et
al. \cite{antonelli}),
was clearly detected in the center of the MECS/LECS field of view. 
The accuracy in the position is $\sim$1'. This accuracy is largely
dominated by uncertainties in the reconstructed BeppoSAX attitude in the
new 1-gyro mode that is implemented since summer 1997.
The S/N ratio for the entire observation is $\sim$12 in the MECS. 
Therefore the source is detected with high significance.
The following data analysis was performed using the SAXDAS data
reduction software, version 1.2, and
the latest release of the LECS and MECS response matrices.

The X-$\gamma$ decay curve is discussed in more detail in Heise et al.
(in preparation). The source faded smoothly during the observation.
A S/N analysis of the count rates accumulated in twenty time intervals
spanning the entire observation shows that the source is visible up to
the end of the NFI observation.
A $\chi^2$ test against constant count rate gives a chance probability
$<10^{-5}$ ($\chi^2_{\rm dof}= 3.7$ for 19 dof).
The same test performed on a light curve extracted in a
source free region is consistent with a constant count rate ($\chi^2_{\rm dof}=
0.76$ for 19 dof).

The spectrum averaged on the entire observing time is consistent with a
single power law with spectral index $\Gamma = 1.6\pm 0.2$ (see Table 1).
The measured value of N$_{\rm H}$ (1$^{+2.3}_{-1}\times 10^{21}$) cm$^{-2}$
is completely consistent with the expected value due to
galactic absorption along the line of sight N$_{\rm H} \approx 1.6\times
10^{20}$ cm$^{-2}$. To have a meaningful upper limit for N$_{\rm H}$
at the GRB frame,
if the association of GRB971214 with the host galaxy is
correct and therefore its redshift is 3.42, the measure of N$_H$ coming
from the formal fit with a non-redshifted function is useless. We
therefore performed also an analysis using a redshifted model. The
power law index is obviously unchanged, while the N$_{\rm H}$ value is
completely not determined. The measured 2-10 keV flux is 2.55$\pm 0.2\times
10^{-13}$\ergs, corresponding to a total 2-10 keV fluence of 4.9$\times
10^{-8}$ erg \rcm2 from 23300 s to 215000 s after the GRB.
In Fig. 3 we show the count rate spectrum with the fitted power law
and the joint confidence contours (68\% and 90\%) for $\Gamma$ and
N$_{\rm H}$.

We analyzed separately the spectrum in three time intervals, to search for
spectral variations. Within the accuracy of our detection, the three
time resolved spectra are consistent with no spectral variation.

\section{The broad band spectrum of the afterglow}

For the afterglow of GRB971214, we collected from the literature (see
caption of Fig. 4 for references) optical and near-IR data taken on 1997
Dec 15, 16 and 17 to construct broad-band spectra from IR to X-rays.  To
determine the magnitudes of the Optical Transient (OT), we first
subtracted from the measurements the contribution of the host galaxy, for
which we considered V=26.5 (Odewahn et al.  \cite{odewahn}),
R=25.6 (Kulkarni et al. \cite{kulkarni_n}), I=25.4, J=25.0, K=24.5,
estimated from H$_{\rm host} = 23.7$ by Fruchter et al. (in preparation)
assuming a flat IR spectrum.  Next, we corrected the resulting OT
magnitudes for the foreground Galactic absorption using A$_{\rm V}$ = 0.1
(from Dickey \& Lockman \cite{dickey}) and the extinction law by Cardelli
et al. (\cite{cardelli}). For each of the three
considered epochs, we referred all data points taken on that night to the
dates 1997 Dec 15.51 (t$_1$), 16.51 (t$_2$), and 17.50 (t$_3$),
respectively. If needed, we rescaled the data to the corresponding
reference date using a power law decay with index $\beta_{\rm t\_opt}$ =
$-$1.2 $\pm$ 0.02 (Diercks et al. \cite{diercks}). 

We also considered a possible dust obscuration at the source redshift,
based on the hypothesis that GRBs could be associated with star formation
(e.g. Paczy\'nski \cite{paczynski}),
and applied to the OT data the extinction law of
Calzetti (\cite{calzetti}) for a typical starburst galaxy at
z=3.42 (details about this correction will be reported in Masetti et
al. in preparation).The
corrected data together with the 2-10 keV fluxes observed at epochs t$_1$
and t$_2$ and the extrapolation at epoch t$_3$ of the X-ray flux in the
same band are reported in Fig. 4. 

\begin{figure}
\epsfxsize=\columnwidth
\epsfbox{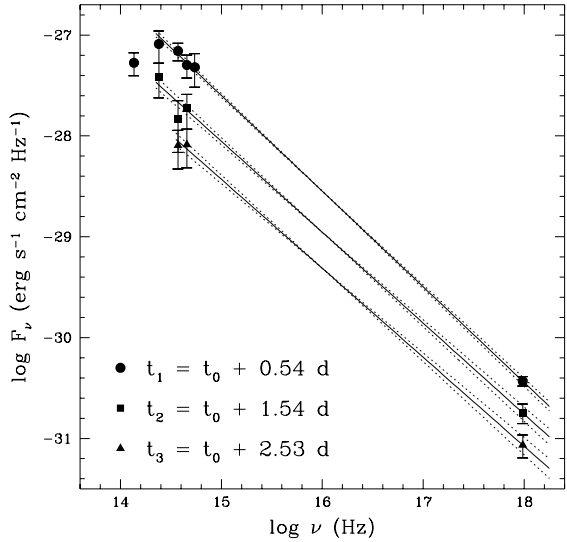}
\caption[]{
Near-infrared, optical and X-ray data of the afterglow of GRB971214 at
three epochs. Optical/IR data are from Diercks et al. (\cite{diercks}),
Halpern et al. (\cite{halpern}), Ramaprakash et al. (\cite{ramaprakash})
and Kulkarni et al. (\cite{kulkarni_n}).
The IR and optical data have been corrected for Galactic and
intrinsic absorption (see text).  The X-ray data are corrected for
Galactic extinction.  The power law fits to the simultaneous
J-band-to-X-ray data are shown (solid lines) along with their 1$\sigma$
confidence ranges (dotted lines). The GRB trigger time is defined as
t$_0$.}
\end{figure}

The optical and X-ray data are well fitted, in a F$_\nu$
versus $\nu$ diagram, by single power laws with slopes  
$\alpha_{\rm opt\_X}$ = $-0.95 \pm 0.02$ (t$_1$), 
$\alpha_{\rm opt\_X}$ = $-0.90 \pm 0.04$ (t$_2$), and 
$\alpha_{\rm opt\_X}$ = $-0.88 \pm 0.04$ (t$_3$). 
These values are all marginally consistent among each other
within the 1$\sigma$ uncertainties
and with the slope of the 2-10 keV X-ray spectrum $\alpha_{\rm X} =
-0.6\pm 0.2$, and are compatible with the value $\alpha_{\rm exp}$ =
$-$0.8, expected for $\beta_{\rm t}$ = $-$1.2 if the peak frequency
$\nu_{\rm m}$ of the afterglow multiwavelength spectrum, produced by a
single synchrotron radiation component, has already passed the optical/IR
band and the cooling frequency $\nu_{\rm c}$ has not ({\it slow cooling
case}, Waxman \cite{waxman_1,waxman_2}, Sari et al. \cite{sari}). 

The presence of a spectral turnover in the IR, already noted by
Ramaprakash et al. (\cite{ramaprakash}), who applied {\it a
posteriori} an extinction correction to the optical/near-IR
afterglow data, could be identified with the change of spectral slope at
the frequency $\nu_{\rm m}$ in the framework of the above-mentioned model
(see however Wijers \& Galama \cite{wijers}).
The rather low statistical quality
of the IR data prevents an accurate measure of the spectral index
$\alpha_{\rm IR} = 0.75^{+1.05}_{-1.15}$ for epoch t$_1$, that is
consistent with the expected value 0.33. 

Afterglow data shown in Fig. 4 are quite remarkable.
The broad-band  spectrum over four decades of
photon energy clearly shows that the  optical and
X-ray emission vary in a coordinated way.
The deduced broad-band energy index
$\alpha'_{\rm opt\_X} \sim -1$ (within uncertainties)
indicates a {\it flat $\nu \, {\rm F}_{\nu}$ spectrum
for three decades of photon energy}.
In a model for which the optical and X-ray  afterglow emission
is produced by a single distribution of energized particles,
the flatness of the $\nu \, {\rm F}_{\nu}$ spectrum  can be
obtained only by a very efficient acceleration process
that has to operate despite the radiation and adiabatic losses.
In our data extending up to 2.5 days after the prompt burst 
emission,
there is no sign of a spectral break due to a transition
between fast and slow cooling. Other bursts behave differently,
with breaks observed both in the lightcurves and spectra
at late times (e.g. GRB990123, Akerlof et al. \cite{rotse},
GRB990510, Harrison et al. \cite{harrison}).

\section{Discussion}

Various authors have discussed the emission in the afterglow (e.g.
Tavani \cite{tavani}, Vietri \cite{vietri}, M\'esz\'aros \&
Rees \cite{meszrees}, Sari et al. \cite{sari}, Waxman 
\cite{waxman_1,waxman_2}), giving a convincing answer to the
problem of light curve modeling.\\
The observed decay during the afterglow (Heise et al. in preparation)
smoothly reconnects
with the observed X-ray flux during the burst. In the hypothesis that
the afterglow starts a few seconds after the end of the main event,
we can
estimate the total fluence in the afterglow, to be compared with
the X-ray luminosity during the prompt event. This hypothesis is
supported by the recent detection of an optical afterglow in GRB990123
a few seconds after the burst trigger (Akerlof et al. \cite{rotse})
and by the interpretation of the observed complex light curve (Sari \&
Piran \cite{sapir_grb99}).
Also the detection of an early power--law--like tail from GRB920723
(Burenin et al. \cite{granat}) supports the hypothesis that the
afterglow starts
immediately after the GRB, and probably without any interruption in the
X--ray flux.
Of course this assumption adds further uncertainty to the total
estimate. As an example, assuming that the afterglow starts $\sim$1000 s
after the trigger of the main event (and not a few seconds after {\it the
end} of the main event) the total integral differs by a factor $\sim$2.

In doing this estimate we have to assume a spectral shape and slope.
Assuming a time decay law $\propto {\rm t}^{-\beta_{\rm t\_X}}$ with
$\beta_{\rm t\_X}\sim$1.2 (Heise et al. in preparation), 
we obtain that the total fluence in the afterglow between 2 and 10 keV
uncorrected for redshift is 4.1 $\times 10^{-7}$ erg \rcm2. It
corresponds to a luminosity of 1.3$\times 10^{52}$ erg.
This is 2 times the total luminosity in the same energy band during the
burst. This is a lower limit, as we do not add any time evolution
of the spectral shape, but we conservatively adopt the measured spectral
index during the afterglow, much steeper than that measured during the
prompt event.

If we assume a similar, or even steeper, spectral slope in the 2-700
keV energy interval (uncorrected for redshift),
the total energy output from the afterglow is
substantial. For the same power law index measured in the 2-10
keV energy band, we obtain a fluence in the afterglow
$~4.5\times 10^{-6} \flux$. Using a more conservative assumption,
adding an exponential cutoff with folding
energy $\sim$50 keV, the total fluence in the afterglow up to 2.5 days
after the main event is 1.15$\times 10^{-6} \flux$.
We conclude that the {\it radiated} X and $\gamma$ total luminosity in 
the afterglow may be estimated to be between 3.6$\times
10^{52}$ ergs and 1.4$\times 10^{53}$ ergs.
This is comparable to the total radiated power in the main GRB event.
This estimate is only speculative, as the spectral data of the
afterglow are consistent with any cutoff energy above $\sim$2 keV
(90\% confidence) but it points out the need for prompt measurements of
the afterglow that can give a good estimate of the spectral shape above
10 keV.\\
Unfortunately the spectral evolution in hard X-rays
(above 10 keV) of most GRBs seems unaccessible to the present generation of
telescopes operating in this energy band. This is a very important
observational point that cannot be fulfilled up to the next generation
of focusing hard X-ray telescopes, maybe at least for a decade.

The time resolved spectral analysis of the afterglow reported in Table 1,
even if of low statistical quality, do
suggest that the spectral shape remains stable during our observation
of the afterglow. While these data cannot be profitably used to
constrain the spectral evolution of the afterglow in X--rays, they
confirm the stability of the {\it engine } that is producing the
observed X--ray emission in spite of the substantial power--law decline
with time in the observed X--ray luminosity.

The spectral evolution from GRB to the X-ray afterglow indicates that
the energy distribution shifts towards lower energies, as expected from
theoretical models. The model for spectral evolution of Sari et
al. (\cite{sari}) suggests that the high energy tail of the emitted
photon spectrum of the afterglow has a power law slope
$\sim-\frac{\rm p}{2} - 1$ in the case of fast cooling,
where p is the index of a power law distribution of the
electrons. Our measurement of a power law slope $\sim -1.6$ implies
p$\sim 1.2$, a value that would give a non-finite energy in the
electrons.
In this case the observations can be reconciled with theory
assuming that a suitable cutoff, e.g. exponential, in the energy
distribution of the parent electron population is present. The
observation of this cutoff in the produced photon distribution is
however beyond the capability of the present generation of X--ray
telescopes and may be accessible to the new missions like {\it XMM} and
{\it Chandra} if it is substantially below 100 keV.
Our data support (as
suggested also by Waxman \cite{waxman_1,waxman_2})
that we observe a regime where the synchrotron cooling time is long
compared to the dynamical time. In this regime
$\Gamma = -\frac{({\rm p}-1)}{2} - 1$, and for a finite power in the electrons
one obtains $\Gamma < -1.5$, definitely compatible with our measurement.
A caveat must be added to this interpretation: as the adiabatic
losses become dominant in this regime, the observed {\it radiated}
luminosity is produced via synchrotron losses in much an inefficient
way. This brings down the efficiency of the radiative process and in
parallel rises accordingly the total energy budget in the afterglow.

The slope of the NIR-to-X-ray afterglow spectrum, corrected for Galactic
and local extinction, and its temporal evolution are in fair agreement
with models of expanding fireballs.  Our assumption on the intrinsic
extinction reasonably conforms with the proposal that GRBs are connected
with star formation, and therefore expected to reside in star forming
regions of their host galaxies (though not necessarily in extreme
starburst galaxies, see Odewahn et al. \cite{odewahn}). We note that the
local extinction correction we adopted has the advantage, with respect
to other approaches, of using a specific model curve and
of making the spectrum consistent with a single radiation component over
more than three decades of frequency.

If the emission is isotropic
the total X-$\gamma$ luminosity in GRB971214, if at redshift z=3.42, is 
quite higher than 10$^{51}$ ergs, as already pointed out by
Kulkarni et al. (\cite{kulkarni_n}). Our analysis shows that
a substantial fraction, more than 60\%, of the total radiated energy
in 2-10 keV is in the afterglow. In addition a reasonable guess of a high
energy exponential cutoff, with a folding energy of 50 keV, brings us to
conclude that the total power in GRB971214 may be grossly underestimated
if based only on the prompt event.

If the efficiency to convert the total energy output from the GRB in the
afterglow is low, e.g. $\sim$10\%, as
usually assumed in most theoretical models of fireball expansion,
our measure of a substantial energy output in 2--10 keV during afterglow 
shows that the total energy balance of a GRB is grossly underestimated.
Furthermore, if we consider the probable presence of a high energy tail
of the afterglow, a conservative estimate may bring
the total energy output from GRB971214 to more than 10$^{54}$
erg for an isotropically emitting source. Alternatively, one may more
comfortably stay with a luminosity of 10$^{53}$ erg assuming a more
efficient mechanism to effectively extract radiative power from the
expanding fireball or assuming an extreme cutoff to the high energy
spectrum.

A way out of this deadlock may be beaming (Yi \cite{yi}, Shaviv \& Dar 
\cite{shdar}).
Different authors have discussed the ``beaming solution'' to the
GRB/afterglow observational problem (Dar \cite{dar1,dar2}, Rhoads
\cite{rhoads}, M\'esz\'aros \& Rees \cite{meszrees}, Drodzova \&
Panchenko
\cite{drodzova}, Panaitescu et al. \cite{panait}). If the emitted
power is strongly beamed, and therefore not isotropically distributed,
the total power in the GRB may be reduced by orders of magnitude, depending
on the beam open angle. Of course this has some major and obvious impacts.
The number of GRBs (not detected at earth) rises by the same orders of
magnitude. Limb darkening (due to the random distribution
of viewing angles inside the emission cone), time dependent (due to
different beaming at different times after the main event) and energy
dependent (due to the time--dependent photon energy distribution) effects
should be observable once the afterglow sample is large enough.
Examples of such effects are discussed in Panaitescu et al.
(\cite{panait}), including a mixed case with a collimated jet and a
contribution from isotropic ejecta.

Assuming a jetlike outflow, the models (Panaitescu et al. \cite{panait},
Rhoads \cite{rhoads2}, Sari et al. \cite{sapir_jets})
predict a steeper decay in the light curve,
depending on the jet opening angle. This steepening compared to an
isotropic fireball expansion occurs for a 10$^{\rm o}$ opening
angle at approximately 6 days after the event, earlier for smaller
angles. We do not detect such a steepening in the X--ray light curve up
to 2.5 days after the main event.
If we assume a beam open angle $\theta \geq 10^{\rm o}$ the total
energy in the event is reduced accordingly, compared to the isotropic case.

An argument to assess jetlike or spherical emission is proposed by
Sari et al. (\cite{sapir_jets} ), using the decay slope of the
high energy afterglow. As reported by other authors (see above), they
suggest that the power law decay for a jetlike emission is appreciably
steeper. For an isotropic fireball the expected decay is
t$^{-\beta_{\rm X}}$ with $\beta_{\rm X} \sim 1.1-1.3$, while for an
expanding jet the decay follows a power law with $\beta_{\rm X} \sim$
2.4 .\\
This effect is purely geometrical, as it is geometrical the effect
of the jet {\it ``spillover'' } (Rhoads \cite{rhoads2}) expanding sideways
at the
local sound speed. In order to maintain the observed time decay power law
$\propto t^{-1.2}$, the Lorentz factor $\gamma$ must be $> \theta^{-1}$
during all the afterglow observation (see e.g. Piran \cite{piran}).
In the case of GRB971214, this must hold up to the last observation of
the power law decay, performed approximately 2.5 days (60 hours)
after the prompt event.
Following Sari et al. (\cite{sapir_jets}) this translates in a lower
limit to the beam opening angle $\theta_0 > 0.1 \times (6.2 \times 60
\times ({\rm E}_{52}/ {\rm n}_1)^\frac{1}{3} )^\frac{3}{8} \approx 0.2$,
assuming the total ``isotropic''
energy in the afterglow is $\sim 10^{53}$ ergs.
This lower limit in beaming angle translates into a lower limit in the
total radiated power from this GRB (prompt+afterglow) $\sim 10^{52}$ ergs.

If the effect of beaming is comparable during the prompt event and the
afterglow, the conclusions we draw on the relative {\it observed} energy
output apply also directly to the total energy balance in the two cases.
If the effect of beaming is evolving from an extremely beamed emission
during the event to a relatively hollow beaming during the part
of the afterglow we observed, the relative energy balance between the
two cases should scale accordingly.
As a consequence, the energy budget in the
afterglow may increase substantially with respect to that in the
prompt event.

In conclusion the measurement of the spectrum during the prompt event
and during the afterglow strongly supports the models for synchrotron
emission from GRB afterglows, with an agreement both in the X-ray band
and in a broader NIR-to-X-rays band (see Fig. 4). The measured
spectral slope is in fair agreement with the requirements of the models
of expanding fireballs, considering also the observed temporal decay. A
simple argument, based on recent models on jetlike emission and on the
expected temporal decay in this case, supports the observation of a
spherical expanding shell or of a moderate beaming ($\theta > 0.2$).
If this is the case, the observed radiated
power in the afterglow is substantial and should be accordingly reproduced
in any model for X-ray afterglow from GRBs.

{\bf Acknowledgements}. 
This research is supported by the Agenzia Spaziale Italiana (ASI) and the
Consiglio Nazionale delle Ricerche (CNR) of Italy. BeppoSAX is a joint
program of ASI and of the Netherlands Agency for Aerospace Programs (NIVR).
We wish to thank all the people of the BeppoSAX Scientific Operation 
Centre and the Operation Control Centre for their skillful and enthusiastic 
contribution to the GRB research program.


\begin{thebibliography}{} 

\bibitem[1999]{rotse}
Akerlof C. W., Balsano R., Barthelmy S., \etal, 1999,  \nature 398, 400
%
\bibitem[1999]{amati}
Amati L., Frontera F., Costa E., \etal,  1999, \aas 138, 403
%
\bibitem[1997]{antonelli}
Antonelli A., Butler R. C., Piro, L., \etal,  1997, \iauc 6792
%
\bibitem[1996]{xspec}
Arnaud K.A. 1996, in Astronomical Data Analysis Software
and Systems V, eds. Jacoby G. H., Barnes J., ASP Conf. Series
volume 101, p. 17 
%
\bibitem[1999]{ot_12}
Bakos G., Sahu K., Menzies J., \etal,  1999, GCN Circular 387
%
\bibitem[1993]{band}
Band D., Matteson J., Ford L., \etal,  1993, \apj , 413, 281
%
\bibitem[1998]{ot_8}
Bloom J. S., Frail D. A., Kulkarni S. R., \etal,  1998, \apjl 508, L21
%
\bibitem[1997a]{boella}
Boella G., Butler R. C., Perola G. C., \etal,  1997a, \aas 122, 299
%
\bibitem[1997b]{boella2}
Boella G., Chiappetti L., Conti G., \etal,  1997b, \aas 122, 327
%
\bibitem[1997]{ot_2}
Bond H. E. 1997, \iauc 6654
%
\bibitem[1999]{granat}
Burenin R. A., Vikhlinin A. A., Gilfanov M. R., \etal,  1999, \aa 344, L53
%
\bibitem[1997]{calzetti}
Calzetti D., 1997, A. J. 113, 162
%
\bibitem[1989]{cardelli}
Cardelli J.A., Clayton G.C., Mathis J.S., 1989, \apj 345, 245
%
\bibitem[1978]{cavallo}
Cavallo G., Rees M., 1978, \mnras 183, 359
%
\bibitem[1997]{costa}
Costa E., Frontera F., Heise J., \etal, 1997, \nature 387, 783
%
\bibitem[1998]{dar1}
Dar A., 1998, \apjl 500, L93
%
\bibitem[1999]{dar2}
Dar A., 1999, \aas 138, 505
%
\bibitem[1990]{dickey}
Dickey J.M., Lockman F.J., 1990, ARA\&A 28, 215
%
\bibitem[1998]{diercks}
Diercks A., Deutsch E.W., Castander F.J., et al., 1998, \apj 503, L105
%
\bibitem[1997]{drodzova}
Drodzova N. D., Panchenko I. E., 1997, \aa 324, L17
%
\bibitem[1997]{ferocigrbm}
Feroci M., Frontera F., Costa E., \etal, 1997, in EUV, X-Ray, and
Gamma-Ray
Instrumentation for Astronomy VIII, eds. Siegmund O. H., Gummin M. A., SPIE
Proceedings 3114, p. 186
%
\bibitem[1997]{frontera}
Frontera F., Costa E., Dal Fiume D., \etal, 1997, \aas 122, 357
%
\bibitem[1999]{frontera_grb}
Frontera F., Amati L., Costa E., \etal, 2000, \apj in press
%
\bibitem[1998]{ot_5}
Galama T., Vreeswijk P. M., Van Paradijs J., \etal, 1998, \nature 395, 670
%
\bibitem[1999]{ot_10}
Galama T., Vreeswijk P. M., Rol E., \etal, 1999, GCN Circular 313
%
\bibitem[1998]{ot_3}
Groot P. J., Galama T. J., Vreeswijk P. M., \etal, 1998, \apjl 502, L123
%
\bibitem[1996]{hakkila}
Hakkila J., Meegan C. A., Horack J. M., \etal, 1996, \apj 462, 125
%
\bibitem[1998]{halpern}
Halpern J. P., Thorstensen J. R., Helfand D. J.,  \etal, 1998,
\nature 393, 41
%
\bibitem[1999]{harrison}
Harrison F. A., Bloom, J. S., Frail, D. A., \etal, 1999, \apjl 523, L121
%
\bibitem[1997]{heise_grb}
Heise J., in't Zand J, Spoliti G., \etal, 1997, \iauc 6787
%
\bibitem[1998]{ot_7}
Hjorth J., Andersen M. I., Pedersen H., \etal, 1998, GCN Circular 109
%
\bibitem[1997]{jager}
Jager R., Mels W. A., Brinkman A. C., \etal, 1997, \aas 125, 557
%
\bibitem[1998]{ot_6}
Jaunsen A. V., Hjorth J., Andersen M. I., \etal, 1998, GCN Circular 78
%
\bibitem[1997]{batse_fluence}
Kippen R. M., Woods P., Connaughton V., et al. , 1997, \iauc 6789
%
\bibitem[1997]{kulkarni_c}
Kulkarni S. R., Adelberger K. L., Bloom J. S., \etal, 1997, GCN Circular 029
%
\bibitem[1998]{kulkarni_n}
Kulkarni S. R., Djorgovski S. G., Ramaprakash A. N., \etal, 1998,
\nature 393, 35
%
\bibitem[1999]{ot_9}
Kulkarni S. R., Djorgovski S. G., Odewahn S. C., \etal, 1999, \nature 398, 389
%
\bibitem[1997]{manzo}
Manzo G., Giarrusso S., Santangelo A., \etal, 1997, \aas 122, 341
%
\bibitem[1997a]{meszrees1}
M\'esz\'aros P., Rees M. J. , 1997a, \apj 476, 232
%
\bibitem[1997b]{meszrees}
M\'esz\'aros P., Rees M. J., 1997b, \apjl, 482, L29
%
\bibitem[1997]{metzger}
Metzger M. R., Djorgovski S. G., Kulkarni S. R., \etal, 1997,
\nature 387, 261
%
\bibitem[1998]{odewahn}
Odewahn S.C., Djorgovski S.G., Kulkarni S.R., et al., 1998, \apj 509, L5
%
\bibitem[1998]{paczynski}
Paczy\'nski B., 1998, \apjl 494, L48
%
\bibitem[1998]{ot_4}
Palazzi E., Pian E., Masetti N., \etal, 1998, \aa 336, L95
%
\bibitem[1999]{ot_11}
Palazzi E., Masetti N., Pian E., \etal, 1999, GCN Circular 377
%
\bibitem[1998]{panait}
Panaitescu A., M\'esz\'aros P., Rees M. J., 1998, \apj 503, 314
%
\bibitem[1997]{parmar}
Parmar A., Martin D. D. E., Bavdaz M., \etal, 1997, \aas 122, 309
%
\bibitem[1995]{piran}
Piran, T. 1995, Proceedings of the Second Huntsville Workshop,
Fishman G. J., Brainerd J. J., Hurley K. eds., AIP Conference Proceedings
307, p. 495
%
\bibitem[1998]{piro}
Piro L., Heise J., Jager R., \etal, 1998, \aa 329, 906
%
\bibitem[1998]{preece}
Preece R. D., Pendleton, G. N., Briggs, M. S., \etal, 1998, \apj 496,
849
%
\bibitem[1998]{ramaprakash}
Ramaprakash A.N., Kulkarni S.R., Frail D.A., et al., 1998, \nature 393, 43
%
\bibitem[1997]{rhoads}
Rhoads J. E., 1997, \apjl 487, L1
%
\bibitem[1999]{rhoads2}
Rhoads J. E., 1999, \apj , submitted ({\it astro-ph/9903399})
%
\bibitem[1999]{sapir_grb99}
Sari R., Piran T., 1999a,\apjl 517, L109
%
\bibitem[1999]{sapir_jets}
Sari R., Piran T., Halpern J. P., 1999b, \apjl 519, L17
%
\bibitem[1998]{sari}
Sari R., Piran T., Narayan R., 1998,\apjl 497, L17
%
\bibitem[1995]{shdar}
Shaviv N. J., Dar A., 1995, \apj 447, 863
%
\bibitem[1997]{tavani}
Tavani M., 1997, \apjl 483, L87
%
\bibitem[1997]{ot_1}
van Paradijs J., Groot P. J., Galama T., \etal, 1997,\nature 386, 686
%
\bibitem[1997]{vietri}
Vietri M., 1997, \apjl 488, L105
%
\bibitem[1997a]{waxman_1}
Waxman E., 1997a, \apjl 485, L5
%
\bibitem[1997b]{waxman_2}
Waxman E., 1997b, \apjl 489, L33
%
\bibitem[1999]{wijers}
Wijers R.A.M.J., Galama T.J., 1999, \apj 523, 177
%
\bibitem[1994]{yi}
Yi I., 1994, \apj 431, 543
%
\end{thebibliography}
\end{document}